\documentclass[10pt, a4paper]{article}

\usepackage[dvipdfmx]{graphicx, color}
\usepackage{wrapfig}
\usepackage{amsmath, amssymb}
\usepackage{bm}
\usepackage{pifont}
\usepackage[dviout]{pict2e}
\usepackage{booktabs}
\usepackage{lscape}
\usepackage{cite}
\usepackage{setspace}

\usepackage{here}

\usepackage{lineno}

\title{Branching ratio measurement of $h \to \mu ^+\mu ^-$ at the ILC}
\author{Shin-ichi Kawada$^{\dagger}$, Jenny List, Mikael Berggren}
\date{}

\begin{document}

\maketitle

\begin{center}
DESY, Notkestra{\ss}e 85, 22607, Hamburg, Germany
\\
$^{\dagger}$ : shin-ichi.kawada@desy.de
\end{center}

\renewcommand{\thefootnote}{\fnsymbol{footnote}}

\begin{abstract}
\footnote[0]{Talk presented at the International Workshop on Future Linear Colliders (LCWS2017), Strasbourg, France, 23-27 October 2017, C17-10-23.2.}
We study the prospects of measurement of the branching ratio of $h \to \mu ^+ \mu ^-$ at the International Linear Collider (ILC).
The study is performed at center-of-mass energies of 250 GeV and 500 GeV, using fully-simulated MC samples with the International Large Detector (ILD) model.
For both center-of-mass energies, the two final states $q\overline{q}h$ and $\nu \overline{\nu}h$ have been analyzed.
For an integrated luminosity of 2000 fb$^{-1}$ at 250 GeV and 4000 fb$^{-1}$ at 500 GeV, corresponding to the H20 running scenario as well as its staged version, the precision on $\sigma \times \mathrm{BR}(h \to \mu ^+ \mu ^-)$ is estimated.
\end{abstract}

\renewcommand{\thefootnote}{\arabic{footnote}}

\section{Introduction}
The investigation of the Higgs boson is one of the most important research topics in recent particle physics.
In the Standard Model (SM), the Yukawa coupling between matter fermions and the Higgs boson is proportional to the fermion's mass.
If we observe any deviations from this proportionality, it is an indication of new physics beyond the SM.

In this study, we focus on the $h \to \mu ^+\mu ^-$ channel.
This is a very challenging analysis because in the SM the branching ratio of $h \to \mu ^+ \mu ^-$ is estimated to be very small: $2.2 \times 10^{-4}$ for the mass of the Higgs boson of 125 GeV~\cite{HiggsBR}.
However this channel is still important, because the mass of the muon has a small uncertainty unlike quarks which typically have large theoretical uncertainties from QCD, which means that this channel will be a suitable probe for the precise measurement.
We can study not only the muon-Yukawa coupling itself, but also the relation between mass and coupling using the coupling ratios of second and third generation leptons ($\kappa _{\mu} / \kappa _{\tau}$), and second generation lepton and quark ($\kappa _{\mu} / \kappa _c$) to understand the mass generation mechanism.

In this study, we estimate the precision expected for the measurement of $\sigma \times \mathrm{BR}(h \to \mu ^+ \mu ^-)$ at the ILC based on full simulation of the ILD detector concept.
Actually, this channel has been studied several times under various settings in linear colliders physics~\cite{mu1, mu2, mu3, mu4, mu5, mu6}, but all studies except Ref.~\cite{mu6} have been performed at a center-of-mass energy ($\sqrt{s}$) of 1 TeV or higher.
In addition, the studies in Refs.~\cite{mu4} and \cite{mu6} are based on the mass of Higgs boson of 120 GeV.
In Ref.~\cite{mu6} for example, the precision of $\sigma \times \mathrm{BR}(h \to \mu ^+ \mu ^-)$ has been estimated to be 91{\%} at $\sqrt{s} = 250$ GeV with 250 fb$^{-1}$, assuming Higgs mass of 120 GeV and Silicon Detector (SiD) concept for the ILC.
In this study, on the other hand, we focus on $\sqrt{s} =$ 250 GeV and 500 GeV, assuming a Higgs mass of 125 GeV for the first time.
This study will give the prospects for measuring this rare decay channel at lower center-of-mass energies.

At the Large Hadron Collider (LHC), the $h \to \mu ^+ \mu ^-$ decay is explored using $pp$ collision data.
The latest results are shown in Ref.~\cite{ATLASmu} by ATLAS and in Ref.~\cite{CMSmu} by CMS.
They also have studied the prospects at the HL-LHC, ATLAS projects $\sim 21{\%}$ precision on the signal strength with 3000 fb$^{-1}$ data~\cite{ATLASHL-LHC}, while the CMS estimate is $\sim 10{\%}$ for the phase-II detector upgrade~\cite{CMSHL-LHC}.
However, all measurements at the LHC are for the cross section times branching ratio $\sigma \times \mathrm{BR}$.
At the ILC on the other hand, most of the measurements are $\sigma \times \mathrm{BR}$, but it is possible to measure the total cross section $\sigma$ itself by using the recoil technique.
By combining $\sigma \times \mathrm{BR}$ and $\sigma$ measurements, we can extract absolute numbers for the branching ratios without model dependencies.
We can also measure the Higgs total width at the ILC, thus we can extract absolute coupling constants~\cite{White}.

\begin{figure}[t]
\centering
\includegraphics[width = 7.0truecm]{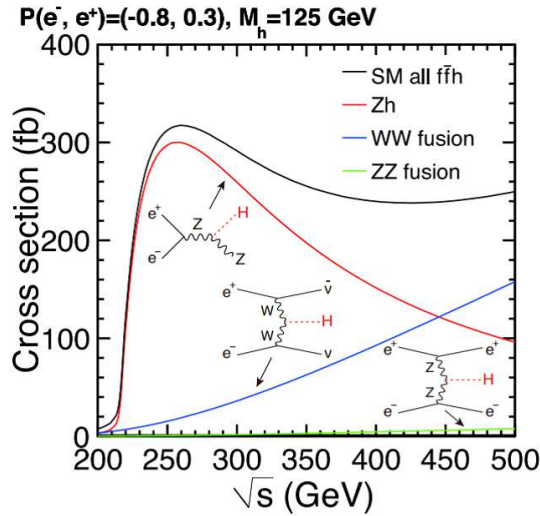}
\caption{The Higgs production cross section as a function of $\sqrt{s}$~\cite{ILCPhysicsCase}.}
\label{fig:xsec}
\end{figure}

The Higgs production cross section as a function of $\sqrt{s}$ at the ILC is shown in Figure~\ref{fig:xsec}, together with corresponding Feynman diagrams.
In this study, we assume the so-called ``H20'' running scenario, accumulating 2000 fb$^{-1}$ at 250 GeV and 4000 fb$^{-1}$ at 500 GeV with actual beam polarization sharing~\cite{ILCPhysicsCase, ILCOperatingScenario}.
The expected number of signal events are summarized in Table~\ref{tab:event}.
We analyze in total 8 channels as listed in Table~\ref{tab:event}.

Recently, the ``staging'' running scenario which starts from 250 GeV operation has been proposed~\cite{250GeVRun}.
We will discuss the prospects with the staging scenario in Section 4.

\begin{table}[t]
\centering
\caption{The expected number of signal events assuming H20 scenario.
The symbols L and R mean the combination of beam polarization of electrons and positrons; L: left-handed, $(e^-, e^+) = (-80{\%}, +30{\%})$, R: right-handed, $(e^-, e^+) = (+80{\%}, -30{\%})$.}
\begin{tabular}{ccc}
\hline
250 GeV & $q\overline{q}h$ & $\nu \overline{\nu} h$ \\
\hline
L & 61.7 (1350 fb$^{-1}$) & 22.5 (1350 fb$^{-1}$) \\
R & 14.1 (450 fb$^{-1}$) & 4.2 (450 fb$^{-1}$) \\
\hline
500 GeV & $q\overline{q}h$ & $\nu \overline{\nu} h$ \\
\hline
L & 24.6 (1600 fb$^{-1}$) & 57.5 (1600 fb$^{-1}$) \\
R & 16.4 (1600 fb$^{-1}$) & 7.9 (1600 fb$^{-1}$) \\
\hline
\end{tabular}
\label{tab:event}
\end{table}

\section{Analysis}
We use fully-simulated Monte-Carlo (MC) samples with the ILD detector model which have been generated in the context of ILC Technical Design Report~\cite{mu2}, with using Whizard~\cite{Whizard} and Pythia~\cite{Pythia}.
We use all available MC samples at 250 GeV and 500 GeV, in total $\sim 8 \times 10^7$ MC events.

The analyzes are structured in the same way in all channels.
First, a pair of well-reconstructed oppositely charged muons consistent with $h \to \mu ^+ \mu ^-$ are selected.
Then, the rest of the event is subject to a procedure to remove the $\gamma \gamma \to$ low $P_t$ hadron overlay and a further, channel-specific selection as a last step of the event selection, a boosted decision tree is applied for each channel.
In this proceedings contribution, we give as an example the details for 500 GeV with $q\overline{q}h$ final state and left-handed beam polarization.
For simplicity, this channel is described as qqh500-L.

\subsection{$h \to \mu ^+ \mu ^-$ Selection}
We apply the so-called IsolatedLeptonTagger~\cite{Tagger} to select $h \to \mu ^+ \mu ^-$ candidate from $e^+ e^- \to q\overline{q}h \to q\overline{q}\mu ^+ \mu ^-$ topology.
In this tagger several variables are used to identify isolated leptons.
For the isolated muon tagging, we require the following conditions: $E_{\mathrm{CAL}} / |p| < 0.5$, $E_{\mathrm{yoke}} > 0.5$ GeV, $|d_0 / \sigma (d_0)| < 5$, $|z_0 / \sigma (z_0)| < 5$, $|p| > 10$ GeV, and MVA cut $> 0.7$, where $E_{\mathrm{CAL}}$ and $E_{\mathrm{yoke}}$ are the energy deposits in the calorimeter and yoke, $p$ is the track momentum, $d_0(z_0)$ is the impact parameter in the $xy(rz)$-plane, $\sigma (d_0) (\sigma (z_0))$ is the measured error of $d_0(z_0)$, respectively.
The final MVA cut is a parameter to check the isolation from other activities.
This tagger can also be used for isolated electrons, but for this muon selection we adjust the variables to make this tagger only behave as the isolated muon tagger.
Thus, the isolated electrons will not be included in the $h \to \mu ^+ \mu ^-$ candidate category.

\begin{figure}[t]
\centering
\includegraphics[width = 7.0truecm]{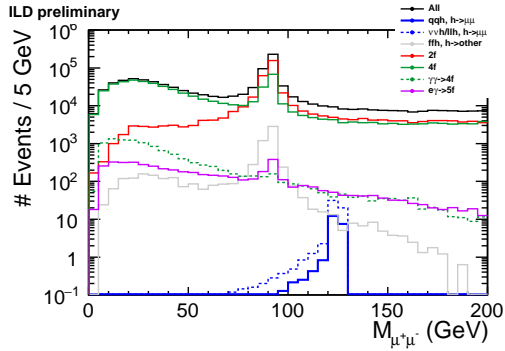}
\caption{The $M_{\mu ^+ \mu ^-}$ distribution before applying the cut to $M_{\mu ^+ \mu ^-}$ (qqh500-L).}
\label{fig:distributions}
\end{figure}

We apply cuts only related to the $h \to \mu ^+ \mu ^-$ candidate as the general event selection.
Since the signal events always have $h \to \mu ^+ \mu ^-$ activities, we can use the general selection as the common cuts for all analyses.
We require following conditions sequentially to $h \to \mu ^+ \mu ^-$ candidate:
\begin{enumerate}
\item exactly one $\mu ^+$ and one $\mu ^-$,
\item $0.5 < \chi ^2 / \mathrm{Ndf} (\mu^{\pm}) < 1.5$,
\item $|d_0(\mu ^{\pm})| < 0.02$ mm、$|d_0(\mu ^-) - d_0(\mu ^+)| < 0.02$ mm,
\item $|z_0(\mu ^{\pm})| < 0.5$ mm、$|z_0(\mu ^-) - z_0(\mu ^+)| < 0.5$ mm,
\item $\sigma (M_{\mu ^+ \mu ^-}) < 1$ GeV for 500 GeV and $< 0.5$ GeV for 250 GeV,
\item $100 < M_{\mu ^+ \mu ^-} < 130$ GeV,
\item $\cos \theta _{\mu ^+ \mu ^-} < 0.55$ for 500 GeV and $< -0.4$ for 250 GeV,
\end{enumerate}
where $\chi ^2 / \mathrm{Ndf}$ is the parameter of how much a track fitted well divided by the number of degrees of freedom of track fit, $\sigma (M_{\mu ^+ \mu ^-})$ is the event-by-event mass resolution, $\theta _{\mu ^+ \mu ^-}$ is the angle between $\mu ^+$ and $\mu ^-$, respectively.
The second and fifth cuts are requiring very well measured tracks and muons, while third and fourth cuts are requiring prompt muons to avoid muons from $\tau$ lepton decay.
The sixth and seventh cuts are used to select only $h \to \mu ^+ \mu ^-$ candidates.
Figure~\ref{fig:distributions} shows the $M_{\mu ^+ \mu ^-}$ spectrum before applying the sixth cut.

\subsection{$Z \to q\overline{q}$}
In the remaining particles after the selection of $h \to \mu ^+ \mu ^-$ candidate, it is expected that it only contains the activities of $Z \to q\overline{q}$ and no isolated leptons.
We again use the IsolatedLeptonTagger~\cite{Tagger} to the remaining particles to count the number of isolated leptons and use for vetoing.
However at 500 GeV, we also have non-negligible contributions from $\gamma \gamma \to$ low $P_t$ hadron overlay~\cite{Overlay}.
To remove these contributions, we use the exclusive $k_T$ clustering algorithm~\cite{kT1, kT2} with a generalized jet radius of 1.0.
We require 4 jets to allow hard gluon emission from each quark.
Any particles not included in these 4 jets are removed since these are most likely coming from $\gamma \gamma \to$ low $P_t$ hadrons background.
After this, we use the Durham clustering algorithm~\cite{Durham} to force the remaining particles into 2 jets, and consider this as the $Z \to q\overline{q}$ candidate.

\begin{figure}[t]
\centering
\includegraphics[width = 7.0truecm]{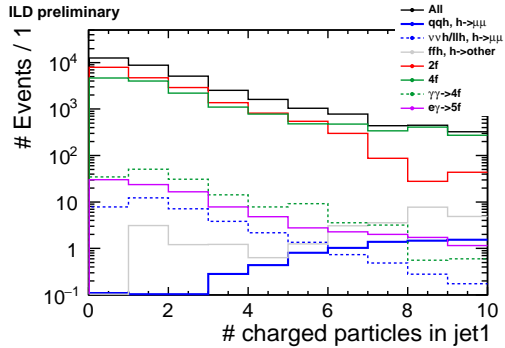}
\caption{The distribution of the number of charged particles in the most energetic jet (qqh500-L).}
\label{fig:distributions2}
\end{figure}

We additionally apply dedicated cuts to select $Z \to q\overline{q}$ candidate and reject background events.
For qqh500-L we apply the following cuts sequentially:
\begin{enumerate}
\item veto: require no isolated leptons in the remaining particles after selecting $h \to \mu ^+ \mu ^-$ candidate,
\item jet clustering successful,
\item after the Durham clustering, each jet should contain at least 4 charged particles,
\item $60 < M_{jj} < 160$ GeV,
\end{enumerate}
where $M_{jj}$ is the invariant mass of the two jets.
The third cut is applied to reject 3-prong $\tau$ decay events, while the fourth cut is selecting $Z \to q\overline{q}$ candidate.
Figure~\ref{fig:distributions2} shows the distribution of number of charged particles in jet1 before applying third cut, where jet1 denotes a jet which has higher jet energy between two jets.
Table~\ref{tab:cuttable} shows the cut table of qqh500-L.

\begin{table}[t]
\centering
\caption{Cut table of qqh500-L.}
\begin{spacing}{1.05}
{\scriptsize
\begin{tabular}{c|ccccccc} \hline
 & $e^+e^- \to q\overline{q}h$ & $e^+e^- \to \nu \overline{\nu} h / \ell \overline{\ell} h$ & $e^+e^- \to f\overline{f}h$ & & & & \\
 & $h \to \mu ^+ \mu ^-$ & $h \to \mu ^+ \mu ^-$ & $h \to$ other & $e^+e^- \to 2f$ & $e^+e^- \to 4f$ & $\gamma \gamma \to 4f$ & $e^{\pm}\gamma \to 5f$ \\
\hline
No cut & 24.6 & 64.1 & $4.12 \times 10^5$ & $4.22 \times 10^7$ & $4.59 \times 10^7$ & $3.36 \times 10^5$ & $2.29 \times 10^5$ \\
\# $\mu ^{\pm}$ & 22.8 & 59.7 & 6455.1 & $1.31 \times 10^6$ & $1.01 \times 10^6$ & $1.49 \times 10^4$ & 5752.4 \\
$\chi ^2$/Ndf & 22.6 & 59.1 & 6396.6 & $1.21 \times 10^6$ & $9.24 \times 10^5$ & $1.31 \times 10^4$ & 5369.9 \\
$d_0$ & 22.5 & 58.8 & 6338.4 & $1.18 \times 10^6$ & $8.51 \times 10^5$ & $1.13 \times 10^4$ & 4978.7 \\
$z_0$ & 22.5 & 58.7 & 6332.1 & $1.17 \times 10^6$ & $8.45 \times 10^5$ & $1.12 \times 10^4$ & 4952.9 \\
$\sigma (M_{\mu ^+ \mu ^-})$ & 22.1 & 58.3 & 6269.1 & $8.03 \times 10^5$ & $8.15 \times 10^5$ & $1.11 \times 10^4$ & 4890.5 \\
$M_{\mu ^+ \mu ^-}$ & 21.5 & 56.6 & 166.0 & $3.83 \times 10^4$ & $2.96 \times 10^4$ & 360.5 & 372.3 \\
$\cos \theta _{\mu ^+ \mu ^-}$ & 21.5 & 56.6 & 121.3 & $2.43 \times 10^4$ & $2.81 \times 10^4$ & 359.9 & 371.5 \\
veto & 21.2 & 52.8 & 115.1 & $2.38 \times 10^4$ & $2.08 \times 10^4$ & 218.5 & 126.4 \\
\# jet & 21.2 & 36.6 & 113.8 & $1.88 \times 10^4$ & $1.68 \times 10^4$ & 159.5 & 101.9 \\
\# charged & 18.4 & 1.5 & 97.7 & 627.4 & 3056.3 & 12.9 & 14.0 \\
$M_{jj}$ & 17.3 & 0.1 & 87.8 & 193.8 & 2298.5 & 4.8 & 9.6 \\
\hline
\end{tabular}
}
\end{spacing}
\label{tab:cuttable}
\end{table}

\begin{figure}[t]
\begin{minipage}{0.48\textwidth}
\centering
\includegraphics[width = 7.0truecm]{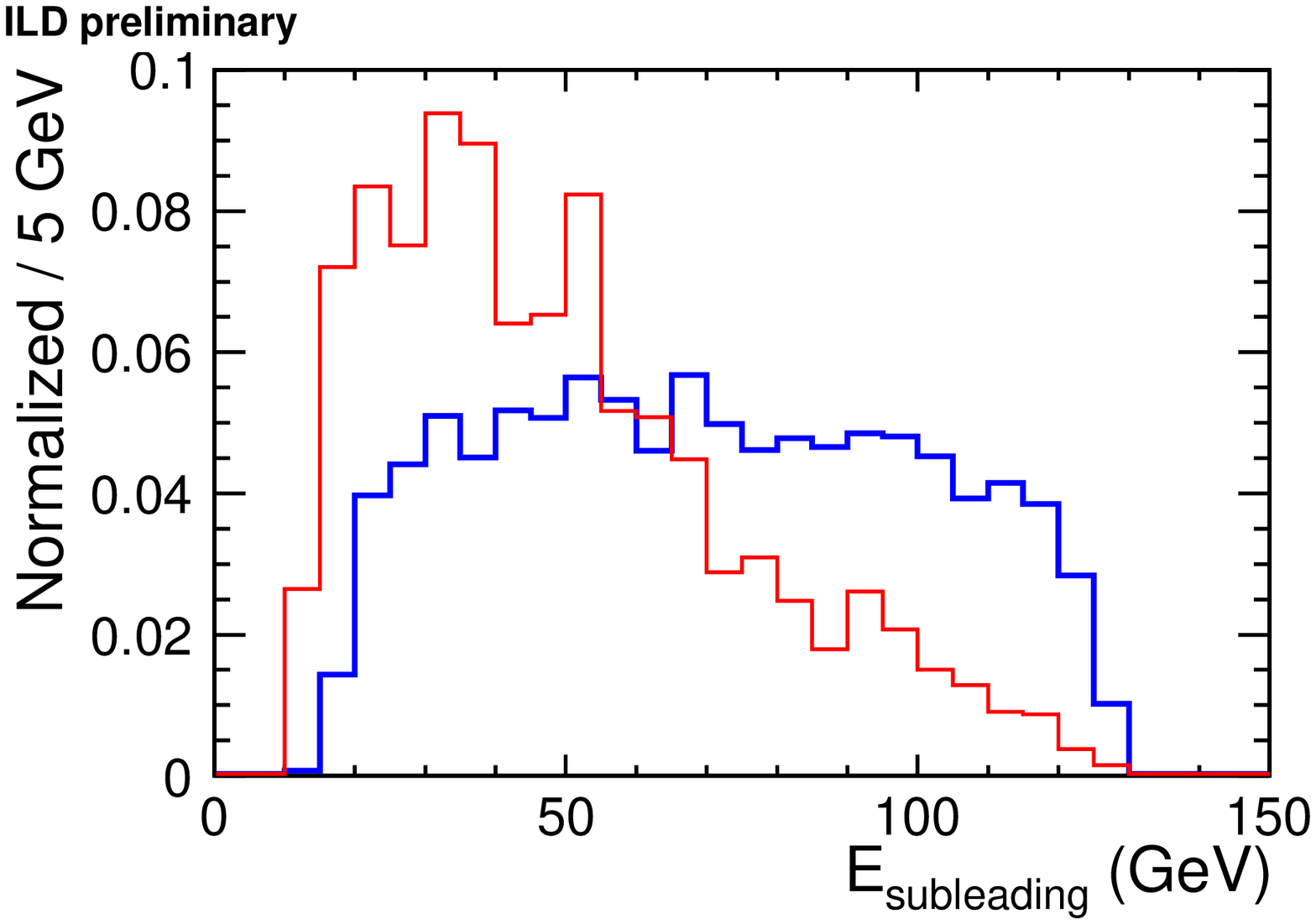}
\caption{Spectrum of $E_{\mathrm{subleading}}$ as the input to BDTG analysis (qqh500-L).
Blue shows signal and red shows background, both histograms are normalized to 1.}
\label{fig:subleading}
\end{minipage}
\begin{minipage}{0.04\textwidth}
\hspace{0.04\textwidth}
\end{minipage}
\begin{minipage}{0.48\textwidth}
\centering
\includegraphics[width = 7.0truecm]{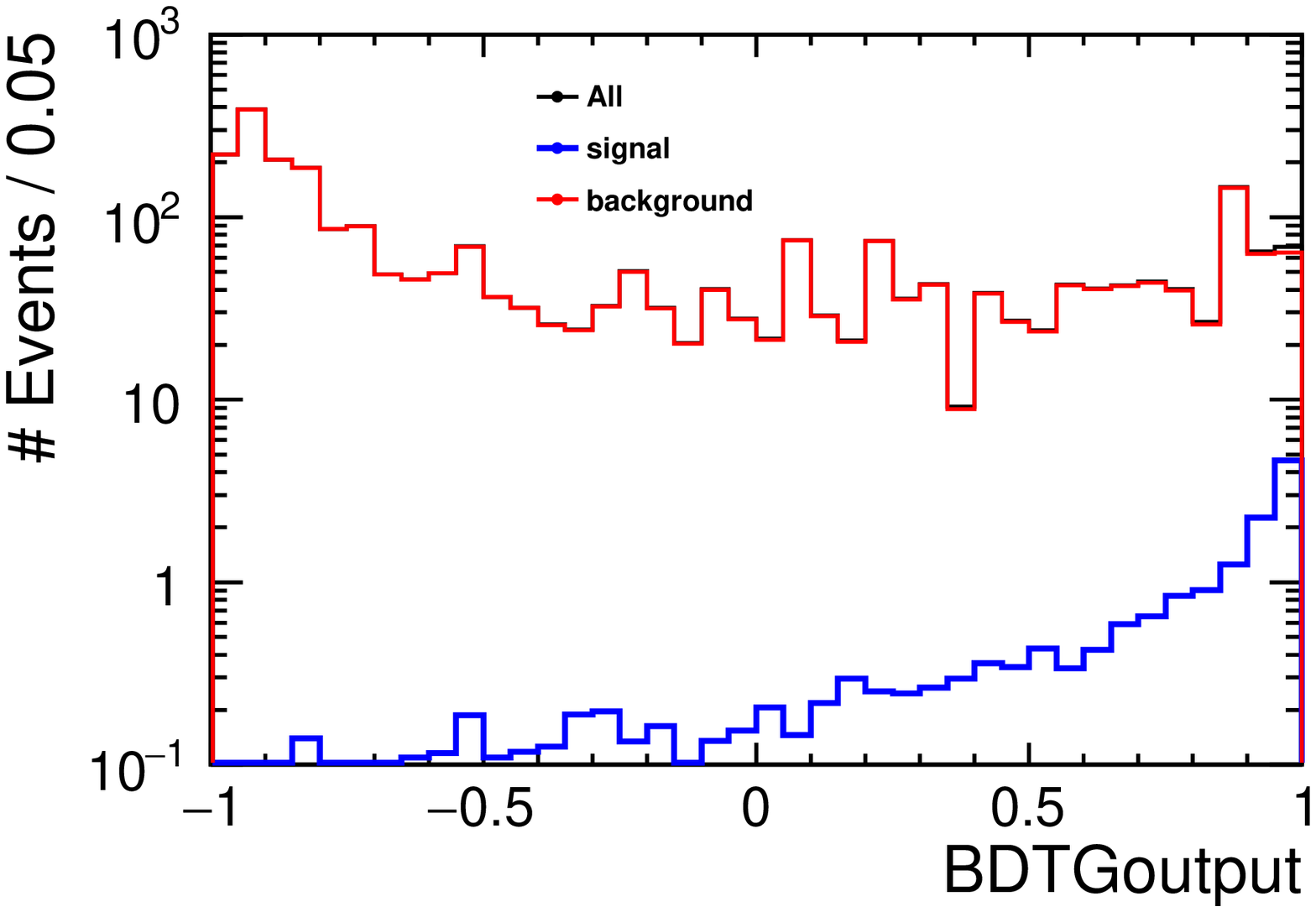}
\caption{Distribution of the BDTGoutput (qqh500-L).}
\label{fig:BDTG}
\end{minipage}
\end{figure}

After all cuts mentioned above, we perform multivariate analysis for further background rejection.
We use gradient boosted decision tree method (BDTG) which is included in TMVA package in ROOT~\cite{TMVA, ROOT}.
For qqh500-L, we use the following 7 variables: thrust, $\cos \theta _h$, charge $\times \cos \theta _{\mu ^+}$, charge $\times \cos \theta _{\mu ^-}$, $E_{\mathrm{leading}}$, $E_{\mathrm{subleading}}$, and $M_{jj}$, where $\theta _h$ is the polar angle of the reconstructed Higgs boson using $h \to \mu ^+ \mu ^-$ candidate, $\theta _{\mu ^+}(\theta _{\mu ^-})$ is the polar angle of $\mu ^+(\mu ^-)$, $E_{\mathrm{leading}}(E_{\mathrm{subleading}})$ is the first(second) largest energy between two muons of $h \to \mu ^+ \mu ^-$ candidate, respectively.
Figure~\ref{fig:subleading} shows the distribution of $E_{\mathrm{subleading}}$ as an example of the input variables to BDTG.
Figure~\ref{fig:BDTG} shows the result of the BDTG analysis.
We apply a cut of BDTGoutput $> 0.65$.
The remaining signal events $N_S$ after this cut are 11.2 while background events $N_B$ are 422.

\section{Results}
\begin{figure}[t]
\begin{minipage}{0.48\textwidth}
\centering
\includegraphics[width = 7.0truecm]{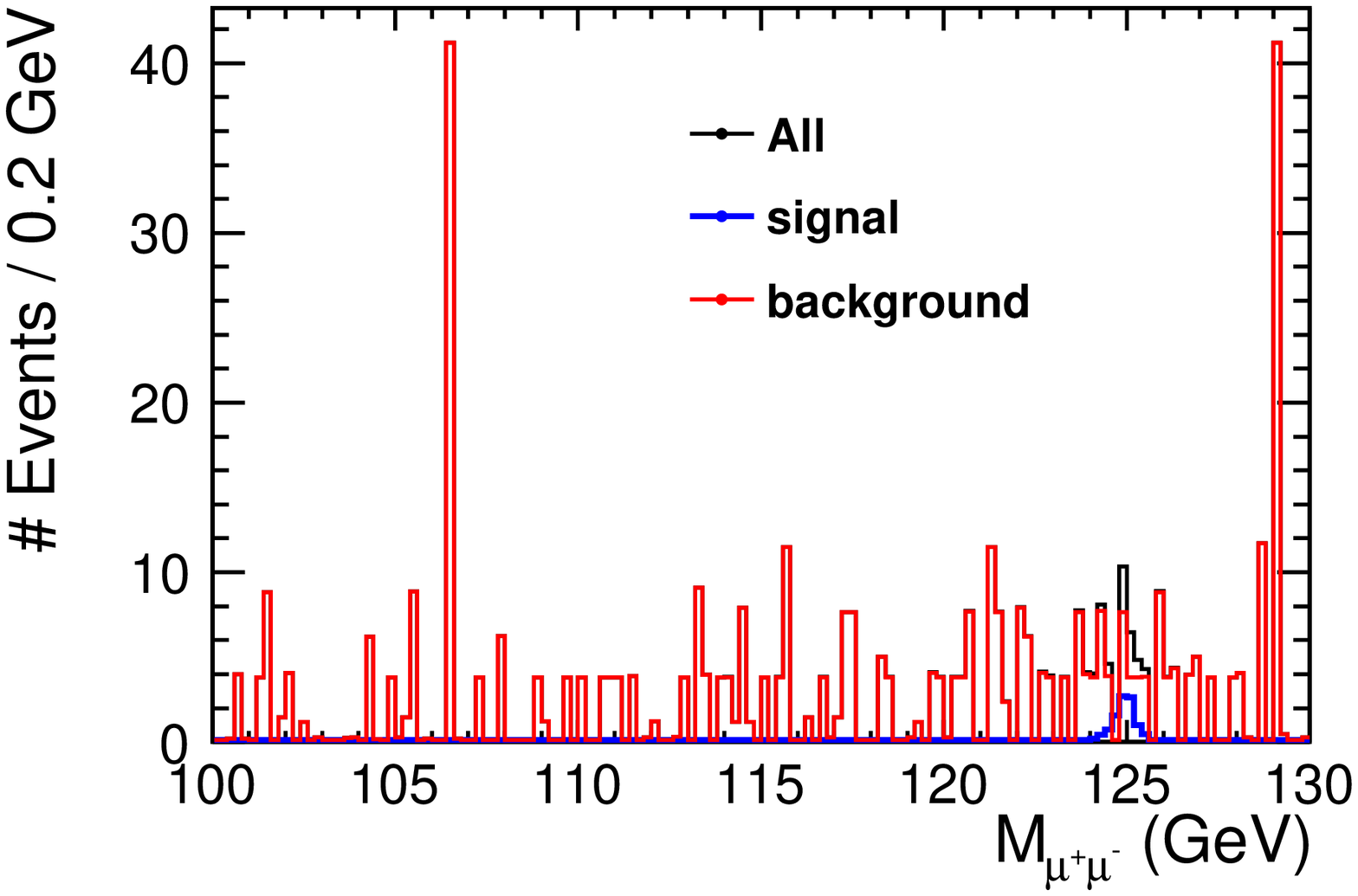}
\caption{Spectrum of $M_{\mu ^+ \mu ^-}$ after all cuts (qqh500-L).}
\label{fig:mumu_mass_final}
\end{minipage}
\begin{minipage}{0.04\textwidth}
\hspace{0.04\textwidth}
\end{minipage}
\begin{minipage}{0.48\textwidth}
\centering
\includegraphics[width = 7.0truecm]{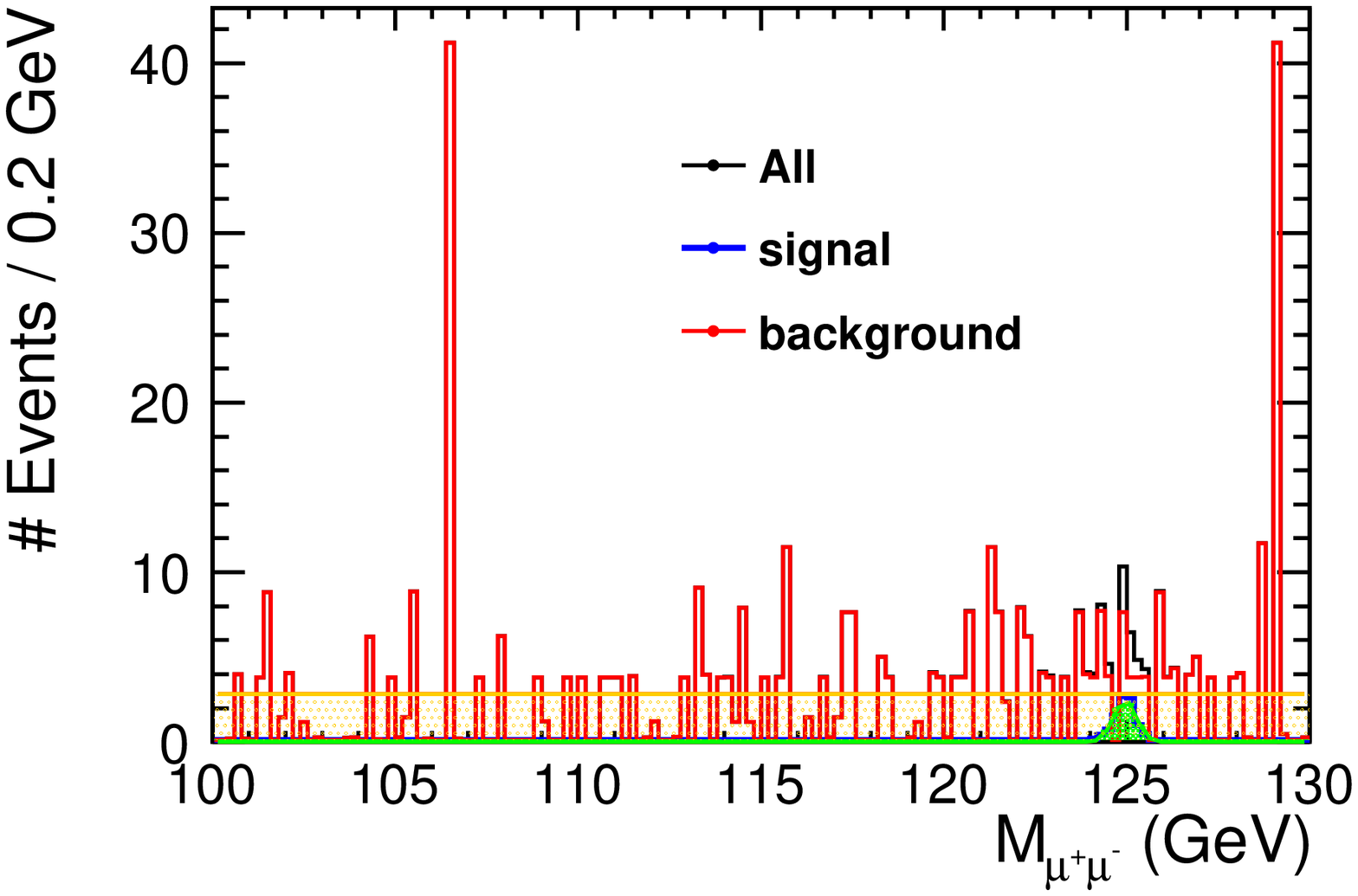}
\caption{Similar to Figure~\ref{fig:mumu_mass_final} but results of the fitting are added.
Green shows the fitting result for signal using $f_S$ and yellow shows for background using $f_B$.}
\label{fig:mumu_mass_fit}
\end{minipage}
\end{figure}

Figure~\ref{fig:mumu_mass_final} shows the $M_{\mu ^+ \mu ^-}$ spectrum after all cuts mentioned in the previous sections.
We can clearly see several spikes in the background distribution, due to the limited MC statistics for SM background.
Therefore, the result will be significantly affected by statistical fluctuations.
To improve this, we apply a toy MC technique to estimate the final uncertainties.

As a first step, analytic functions are fitted to the relevant signal and background histograms.
We use a normalized Gaussian as the signal fitting function $f_S$ and a constant as the background fitting function $f_B$.
Figure~\ref{fig:mumu_mass_fit} shows the result of fitting to $M_{\mu ^+ \mu ^-}$ spectrum using $f_S$ and $f_B$.

\begin{figure}[t]
\centering
\includegraphics[width = 7.0truecm]{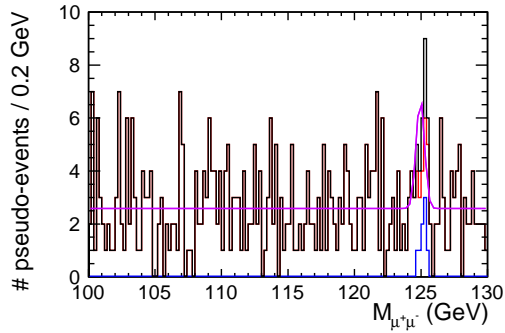}
\caption{One example of pseudo-experiment (qqh500-L).
Blue(red) shows pseudo-signal(background) data, black is the sum of blue and red, purple is the fitting results to black using function $f$, respectively.}
\label{fig:pseudo}
\end{figure}

Then we perform pseudo-experiments using the obtained $f_S$ and $f_B$.
In one pseudo-experiment, the number of pseudo-events are determined by $N_S(N_B)$ with Poisson fluctuation.
Figure~\ref{fig:pseudo} shows an example of one pseudo-experiment.
We use the function $f \equiv Y_Sf_S + Y_Bf_B$ as the fitting function where $Y_S$ is the signal event yield, $Y_B$ is the background event yield, and both are free parameters in the fit.
The purple curve in Figure~\ref{fig:pseudo} shows the result of fitting using $f$ for sum of the pseudo-data.

We repeat pseudo-experiments for 200000 times and obtain $Y_S$ distribution and pull distribution.
From the Gaussian fit to the $Y_S$ distribution, we obtain the mean value of $10.93 \pm 0.01$ and the width of $5.227 \pm 0.008$.
The resulting precision for $\sigma \times \mathrm{BR}(h \to \mu ^+ \mu ^-)$ is estimated to be $47.8{\%}$.
The pull is defined as $(Y_S - Y_{\mathrm{true}}) / \Delta Y_S$, where $\Delta Y_S$ is the fitting error of $Y_S$ and $Y_{\mathrm{true}}$ is corresponding to the number of pseudo-data determined as $N_S$ with Poisson fluctuation.
If there are no biases in the fitting, the pull distribution should have the mean of $\sim 0$ and width of $\sim 1$.
However, we obtain the mean of $-0.071 \pm 0.002$ and width of $0.779 \pm 0.001$ from Gaussian fitting to pull distribution.
This result indicates that there are some biases included in our analysis.
In addition, we find asymmetric distribution for the $Y_S$ and pull distribution.
The reason is under investigation.

In a similar way, we have analyzed all channels listed in Table~\ref{tab:event}.
The results are summarized in Table~\ref{tab:results}.
By combining all 250 GeV results, we can obtain 25.0{\%} combined precision on the cross section times branching ratio $\sigma \times \mathrm{BR}$.
This result is much better than SiD results~\cite{mu6} with statistical scaling extrapolation ($\sim 39{\%}$ for left-handed 250 GeV $q\overline{q}h$ and $\nu \overline{\nu} h$ channels).
Together with the 500 GeV results, the combined precision is estimated to be 17.5{\%}.
This is comparable to ATLAS HL-LHC prospects~\cite{ATLASHL-LHC}, but worse than CMS HL-LHC prospects~\cite{CMSHL-LHC} due to the statistics of number of signal events.
However as we explained in Section 1, we can extract absolute couplings together with other measurements at the ILC without model dependencies, while LHC results always have model dependencies.

\begin{table}[t]
\centering
\caption{Summary of the precision of $\sigma \times \mathrm{BR}(h \to \mu ^+ \mu ^-)$.}
\begin{tabular}{ccc} \hline
250 GeV & $q\overline{q}h$ & $\nu \overline{\nu} h$ \\
\hline
L & 30.0{\%} & 123.5{\%} \\
R & 52.5{\%} & 125.4{\%} \\
\hline
500 GeV & $q\overline{q}h$ & $\nu \overline{\nu} h$ \\
\hline
L & 47.8{\%} & 39.2{\%} \\
R & 52.1{\%} & 71.5{\%} \\
\hline
\end{tabular}
\label{tab:results}
\end{table}

\section{Further Study}
After the LCWS2017, we have studied the case of staging scenario~\cite{250GeVRun}, and investigated further improvements.
In the staging scenario, the beam polarization sharing for 250 GeV is changed from H20 scenario~\cite{250GeVRun}.
The expected number of signal events are summarized in Table~\ref{tab:event250GeV}.

\begin{table}[t]
\begin{minipage}{0.48\textwidth}
\centering
\caption{The expected number of signal events at $\sqrt{s} = 250$ GeV assuming the staging scenario.}
\begin{tabular}{ccc}
\hline
 & $q\overline{q}h$ & $\nu \overline{\nu} h$ \\
\hline
L & 41.1 (900 fb$^{-1}$) & 15.0 (900 fb$^{-1}$) \\
R & 28.1 (900 fb$^{-1}$) & 8.4 (900 fb$^{-1}$) \\
\hline
\end{tabular}
\label{tab:event250GeV}
\end{minipage}
\begin{minipage}{0.04\textwidth}
\hspace{0.04\textwidth}
\end{minipage}
\begin{minipage}{0.48\textwidth}
\centering
\caption{Summary of the precision of $\sigma \times \mathrm{BR}(h \to \mu ^+ \mu ^-)$ with further study and staging scenario.}
\begin{tabular}{ccc} \hline
250 GeV & $q\overline{q}h$ & $\nu \overline{\nu} h$ \\
\hline
L & 32.5{\%} & 108.6{\%} \\
R & 28.1{\%} & 110.4{\%} \\
\hline
500 GeV & $q\overline{q}h$ & $\nu \overline{\nu} h$ \\
\hline
L & 44.5{\%} & 37.0{\%} \\
R & 49.5{\%} & 74.5{\%} \\
\hline
\end{tabular}
\label{tab:results_further}
\end{minipage}
\end{table}

We apply the same analysis procedure except the optimization of BDTGoutput cut and the way of toy MC.
The BDTGoutput cut mentioned at the end of Section 2 was not optimized.
We have studied the optimum cut on BDTGoutput together with the result of the toy MC procedure, and adopted the best case as the optimum result.
Furthermore, the function $f = Y_Sf_S + Y_Bf_B$ using the fitting to pseudo-data, we fix $Y_B$ as $N_B$ which is the number of remaining background after BDTGoutput cut.
Since we have found that $Y_B$ can be determined more precisely compare to $Y_S$ and its precision is $\sim 5{\%}$ or better, it is possible to fix $Y_B$.

The new results are summarized in Table~\ref{tab:results_further}.
The combined precision for 250 GeV is estimated to be 20.5{\%}, and all combined result is 15.4{\%}.
The combined 250 GeV result is relatively $\sim 20{\%}$ improved from the result in Section 3.
The all combined result is also relatively $\sim 10{\%}$ improved.

In summary, we studied the prospects of the branching ratio measurement of $h \to \mu ^+ \mu ^-$ at the ILC assuming ILD detector model in the H20 running scenario as well as in its staged version.
The combined precision using all 250 GeV results is estimated to be 20.5{\%} for $\sigma \times \mathrm{BR}(h \to \mu ^+ \mu ^-)$, which presents a considerable improvement with respect to a previous study at 250 GeV.
Together with the 500 GeV results, the combined precision improves to 15.4{\%}, which is similar to the HL-LHC prospects.
We are planning to analyze $e^+ e^- \to \ell ^+ \ell ^- h$ channel, and work on more background rejection for further improvement.

\section*{Acknowledgements}
We would like to thank the LCC generator working group and the ILD software working group for providing the simulation and reconstruction tools and producing the Monte Carlo samples used in this study.
This work has benefited from computing services provided by the ILC Virtual Organization, supported by the national resource providers of the EGI Federation and the Open Science GRID.
We thankfully acknowledge the support by the Deutsche Forschungsgemeinschaft (DFG) through the Collaborative Research Centre SFB 676 Particles, Strings and the Early Universe, project B1.

\end{document}